\documentclass[aps, preprint, prb,superscriptaddress,showpacs]{revtex4-1}
\usepackage{graphicx}
\usepackage{upgreek}
\usepackage{amssymb,amsmath}

\newcommand{\sto}{SrTiO$_3$}
\newcommand{\ngo}{NdGaO$_3$}
\newcommand{\smo}{SrMnO$_3$}
\newcommand{\lmo}{LaMnO$_3$}
\newcommand{\lsmo}{La$_{0.67}$Sr$_{0.33}$MnO$_3$}

\begin{document}
\title{Interface Engineering in La$_{0.67}$Sr$_{0.33}$MnO$_3$--SrTiO$_3$ Heterostructures}
\author{Hans Boschker}
\email{h.boschker@fkf.mpg.de}
\affiliation{Max Planck Institute for Solid State Research, 70569 Stuttgart, Germany}
\affiliation{Mesaplus Insititute for Nanotechnology and Faculty of Science and Technology, University of Twente, 7500AE Enschede, The Netherlands}
\author{Zhaoliang Liao}
\author{Mark Huijben}
\author{Gertjan Koster}
\author{Guus Rijnders}
\affiliation{Mesaplus Insititute for Nanotechnology and Faculty of Science and Technology, University of Twente, 7500AE Enschede, The Netherlands}

%\pacs{77.65.-j,77.80.Dj,61.05.cp}
%\date{\today}

\begin{abstract}
Interface engineering is an extremely useful tool for systematically investigating materials and the various ways materials interact with each other. We describe different interface engineering strategies designed to reveal the origin of the electric and magnetic dead-layer at \lsmo~interfaces. \lsmo~is a key example of a strongly correlated peroskite oxide material in which a subtle balance of competing interactions gives rise to a ferromagnetic metallic groundstate. This balance, however, is easily disrupted at interfaces. We systematically vary the dopant profile, the disorder and the oxygen octahedra rotations at the interface to investigate which mechanism is responsible for the dead layer. We find that the magnetic dead layer can be completely eliminated by compositional interface engineering such that the polar discontinuity at the interface is removed. This, however, leaves the electrical dead-layer largely intact. We find that deformations in the oxygen octahedra network at the interface are the dominant cause for the electrical dead layer.

\end{abstract}

\maketitle

\section{Introduction}
Oxide epitaxy has made tremendous advances in the past 15 years. Using techniques such as pulsed laser deposition (PLD) and molecular beam epitaxy (MBE), oxide epitaxy has evolved from merely depositing material with the right crystal structure onto a substrate to growing the material in a layer-by-layer fashion. The latter allows interventions during the growth and therefore enables the creation of artificial materials. The progress in epitaxy is due not only to improvements in the growth systems and growth monitoring, but also to a better understanding of the ways materials influence each other during heteroepitaxy. Research has shown that perovskite oxide epitaxial layers are influenced by the boundary conditions applied by the epitaxy process. Properties are modified due to the presence of such phenomena as strain, polar discontinuities, and octahedral connectivity mismatch. This sensitivity offers the unique opportunity to engineer the materials’ structure. In this way, advanced epitaxy has become a research tool for understanding materials, because it allows different structural parameters to be disentangled. Some examples of the possibilities of advanced epitaxy are: tricolor superlattices \cite{Lee2005}, compositional interface engineering \cite{Boschker2012, Peng2014, Huijben2015}, modulation doping \cite{Chen2015}, oxygen octahedra rotation engineering \cite{Bousquet2008, Boschker2012prl, Rondinelli2012, Aso2014, Moon2014, Zhai2014}, interface dipole engineering \cite{Yajima2011}, transfer of electron-phonon coupling \cite{Driza2012}, and defect mitigation \cite{Lee2013}. 

\lsmo~is interesting because it is a ferromagnetic metal with 100 \% spin polarization \cite{Park1998}. It is therefore relevant to a variety of devices such as magnetic tunnel juctions \cite{Sun1999, Bowen2003}, diodes \cite{Postma2004, Hikita2009}, transistors \cite{Yajima2011}, and ferroelectric tunnel junctions \cite{Garcia2010}. All of these applications require control of the properties of \lsmo~at the interface. The half-metallic state in \lsmo~arises from a complex balance between competing interactions. The double exchange interaction favors conductivity whereas the superexchange interaction favors an insulating state. Furthermore, the material is susceptible to Jahn-Teller distortions that also favor electron localization. All these factors depend on the local crystal environment of the Mn ions. At interfaces the balance between the competing interactions is modified and an electrical and magnetic dead-layer is present \cite{Kavich2007, Huijben2008}. 

Here we present a study of different interface engineering schemes for \lsmo~interfaces with a focus on the \lsmo--\sto~interface. The goal is to systematically investigate different mechanisms thought to be responsible for the dead-layer behavior. In the following sections we discuss the polar discontinuity problem, the effect of disorder, and the oxygen octahedra rotations.

\section{Experimental}
\sto~substrates were prepared using a standard BHF etching procedure \cite{Koster1998}.
The \lsmo~thin films were grown with pulsed laser deposition (PLD) (TSST system). The substrate temperature during growth was 750-800 $^{\circ}$C in an oxygen environment of 0.27 mbar. The laser beam was produced by a 248-nm-wavelength KrF excimer laser (LPXPro$^\textrm{TM}$ from Coherent, Inc.) with a typical pulse duration of 20-30 ns. With a 4 by 15 mm rectangular mask the most homogeneous part of the laser beam was selected. A sharp image of the mask was created on the stoichiometric target (Praxair electronics) with a lens, resulting in a spotsize of 2.3 mm$^2$ (0.9 by 2.5 mm). The beam energy was controlled with a variable attenuator or with the laser voltage, yielding a fluence at the target of 2 J/cm$^2$. The repetition rate was 1 Hz and the substrate was placed at 5 cm distance directly opposite to the target. Before deposition, the target was pre-ablated for 2 minutes at 5 Hz to remove any possible surface contamination. After deposition, the PLD chamber was flooded with pure oxygen (typically 100 mbar) and the samples were cooled down by switching of the heater power. Typically, the cooldown required 2 hours. For the PLD growth of \sto~and \smo, identical settings were used as for the \lsmo~growth. For the LaTiO$_3$ growth, and the growth of the \sto~layer on top of the LaTiO$_3$, the deposition pressure was lowered to 3$\cdot10^{-5}$ mbar, in order to prevent the formation of La$_2$Ti$_2$O$_7$ \cite{Ohtomo2002}. Some of the thicker \lsmo~films were grown with a repetition rate of 5 Hz and an oxygen pressure of 0.16 mbar. 

The magnetization of the samples was measured with a vibrating sample magnetometer (VSM) (Physical Properties Measurement System (PPMS) by Quantum Design). In order to determine the saturation magnetization and Curie temperature unambiguously, magnetic hysteresis loops were obtained at all temperatures as discussed in detail in the supplementary information of \cite{Boschker2012}. The resistivity of the samples was measured in the van der Pauw configuration \cite{vdPauw1958} (PPMS by Quantum Design). In order to obtain ohmic contacts between the aluminium bonding wires and the \lsmo~layer, gold contacts were deposited on the corners of the sample with the use of a shadow mask. The electric field gating experiments were performed with a backgate geometry, using the 0.5 mm substrate as a gate dielectric. In order to ensure a uniform electric field, the back of the substrate was glued to a copper plate using silver epoxy. 

\section{Thin film growth and characterization}

\begin{figure}[!htbp]
\centering
\includegraphics*[width=8cm]{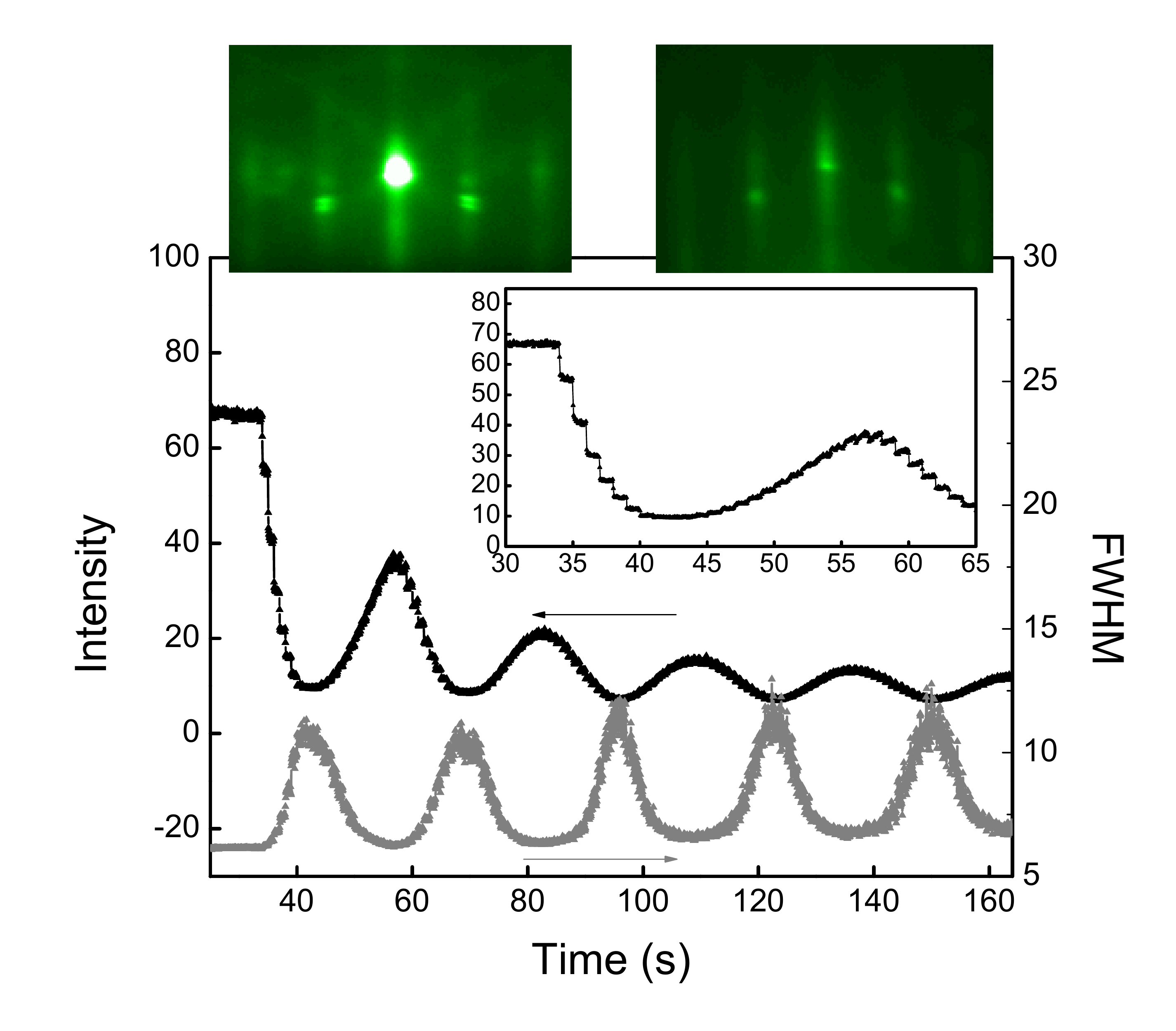}
\caption{RHEED specular spot intensity and FWHM as monitored during the initial growth of \lsmo~on \sto~(001)$_\textrm{c}$. The inset shows an expanded view of the growth of the first unit cell layer. Two RHEED images are presented, the left image was taken before deposition (at low pressure) and the right image was taken after deposition of 5 unit cell layers. Figure reprinted from \cite{Boschker2011}.}
\label{rheedini}
\end{figure}

\begin{figure}[!htbp]
\centering
\includegraphics*[width=8cm]{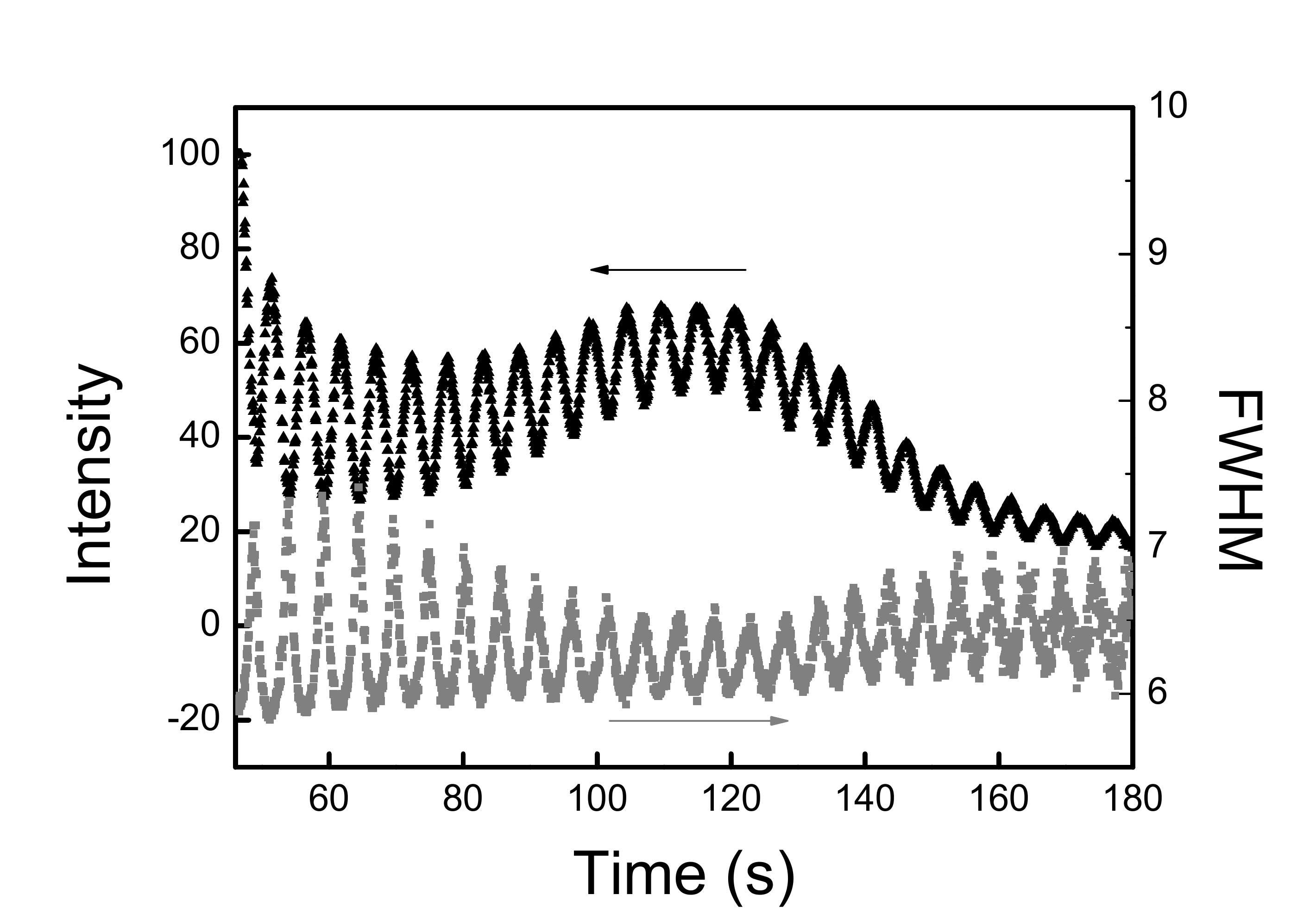}
\caption{RHEED specular spot intensity and FWHM as monitored during the growth of a relatively thick film of \lsmo~on \sto~(001)$_\textrm{c}$. The graph shows the 7$^\textrm{th}$ to 32$^\textrm{nd}$ oscillation. A recovery of the RHEED oscillation intensity maximum, followed by a rapid decrease, is observed around the 20$^\textrm{th}$ unit cell layer. Figure reprinted from \cite{Boschker2011}.}
\label{rheed25}
\end{figure}

The growth of the films was studied \textit{in situ} with RHEED during the growth. The substrate RHEED pattern is shown in Fig.~\ref{rheedini} on the top left. The main specular spot is very intense compared to the two side spots. This is the typical signature of TiO$_2$ terminated \sto~\cite{Koster2000}. Kikuchi lines are visible as well, indicating the smoothness of the substrate. The side spots are doubled, which is due to the additional periodicity at the surface from the regularly spaced terrace steps. Figure~\ref{rheedini}, top right, shows the RHEED image of the \lsmo~film after the deposition of 5 unit cell layers. Clear two dimensional spots are visible, but also 2D streaks are present. The latter are attributed to the scattering of the RHEED beam off the unit cell high steps at the surface. Similar RHEED images were observed after the completion of films with thicknesses up to 40 nm. 

The main graph in Fig.~\ref{rheedini} shows the intensity of the specular reflection as measured during the initial growth of \lsmo. The intensity shows oscillations which correspond to the growth of the individual unit cell layers. Within the oscillations, recovery of the intensity after the sudden decrease during the laser burst is observed, as shown more clearly in the inset of the graph. The oscillation amplitude decreases with the amount of material deposited during the first part of the growth. This decrease of the intensity is due to the difference in reflectivity of the \sto~surface and the \lsmo~surface, due to the increased scattering from the heavy La ions in the lattice. Finally, the full width at the half maximum (FWHM) of the intensity of the specular spot, measured along the (10) direction, is presented as well. During the growth, the RHEED spots are periodically more streaky, because more stepedge scattering is present at half unit cell layer coverage compared to full unit cell layer coverage. Therefore, the FWHM oscillates as well. The FWHM depends only on the shape of the intensity distribution and not on the total intensity and it is therefore a better indicator of the surface morphology than the main intensity of the reflection. As can be seen in the figure, the FWHM during the growth is almost equal to the FWHM of the substrate reflection indicating a smooth surface morphology. From these measurements it is concluded that the initial stage of the \lsmo~growth proceeds in the ideal 2D layer-by-layer growth mode.   

During the growth of the film, an increase of the RHEED oscillation maximum intensity was observed, as presented in Fig.~\ref{rheed25}. The oscillation intensity maximum typically peaked around 20 to 25 unit cell layers. The maximum in intensity was not observed during growth on an $A$-site terminated substrate surface, so it might indicate a termination conversion. After this peak the oscillation intensity maximum decreased rapidly and the oscillation amplitude became comparable to the intensity variations in the laser pulse recovery cycles. It is concluded that the \lsmo~growth mode during the latter part of the growth is close to the steady state growth mode which is characterized by a relatively constant step density, which is large compared to the step density of the initial substrate surface. 

\begin{figure}
\centering
\includegraphics*[width=10cm]{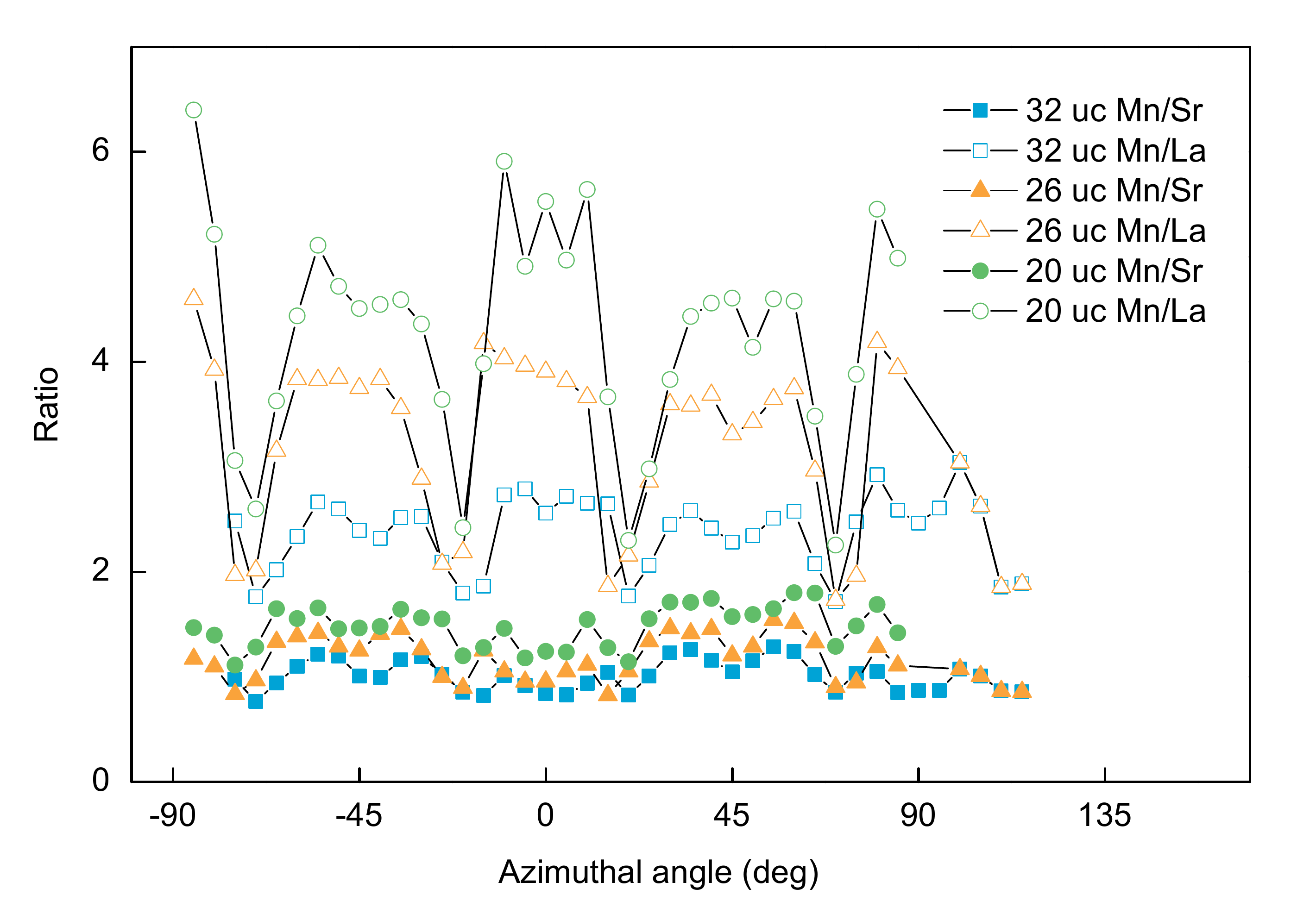}
\caption{Angle-dependent time-of-flight mass spectroscopy. The angular dependence of the relative intensities is due to the atomic shadowing effect.}
\label{tof}
\end{figure}

In order to check the hypothesis about a termination conversion during growth, three samples with respectively a 20, 26 and 30 unit-cell-thick \lsmo~layer were measured with angle-dependent time-of-flight mass spectroscopy. The relative intensity of the Mn ions with respect to Sr and La is shown in Fig.~\ref{tof}. The strong angular dependence in the Mn/La ratio of the 20 unit-cell-thick sample is consistent with a dominant MnO$_2$ crystal termination. With increasing film thickness the angular dependence is reduced indicating a reduction of the MnO$_2$ termination. The data is more consistent with a mixed termination for the thicker films than with a complete surface termination conversion. The Mn/Sr ratio has a similar angular dependence, although the magnitude of the signal is reduced. The lower Mn/Sr ratio compared to Mn/La is also seen in experiments using controlled $A$O and $B$O$_2$ terminated \lsmo~films \cite{Yu2012}. In conclusion, the surface termination changes around 26 unit cells towards mixed termination, in agreement with the deductions from the growth analysis. 

\begin{figure}[!htbp]
\centering
\includegraphics*[width=10cm]{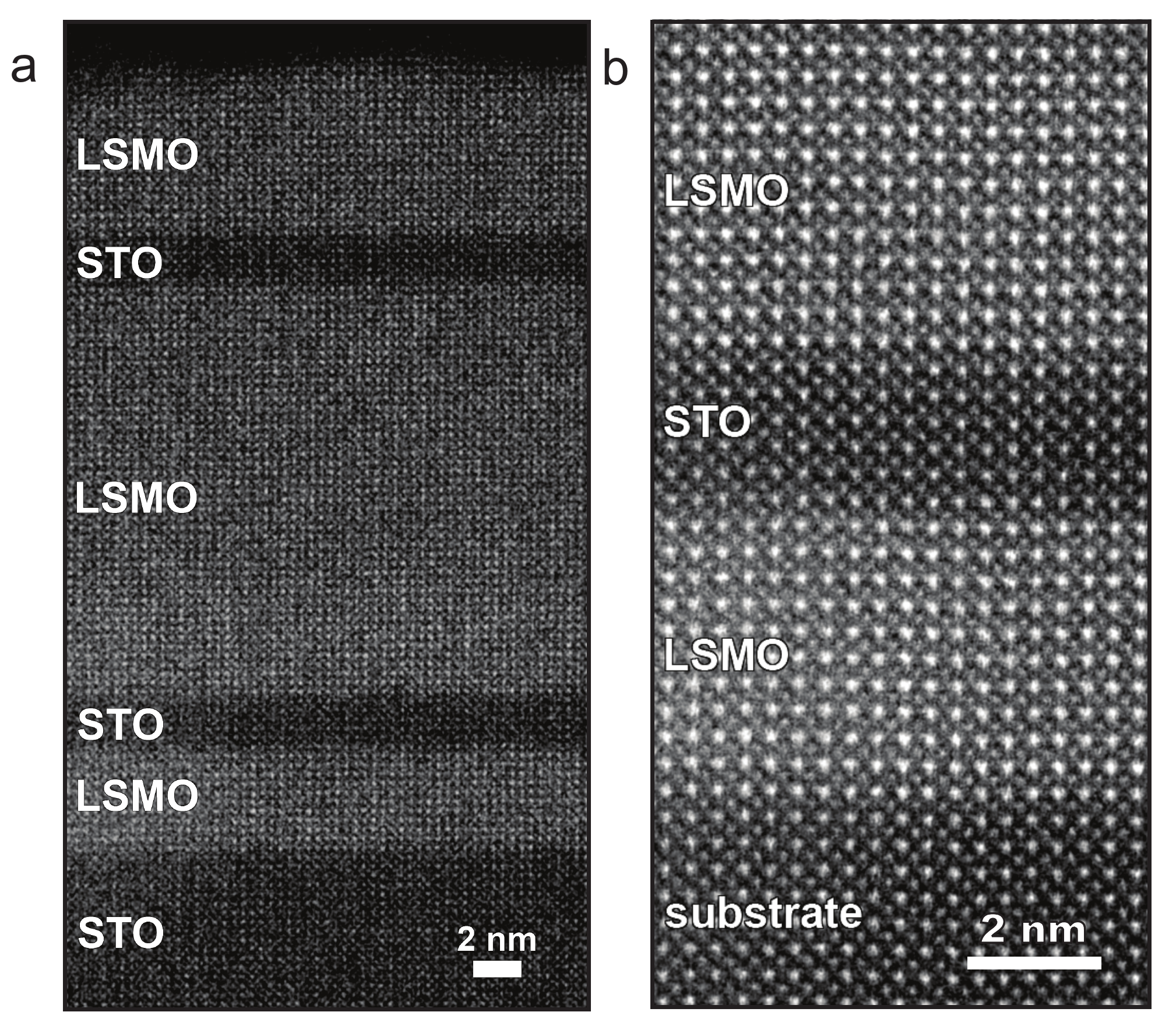}
\caption{Scanning transmission electron microscopy. a) HAADF-STEM image of a \lsmo--\sto heterostructure. b) High-magnification image of the area close to the substrate. Figure reprinted from \cite{Boschker2011}. }
\label{stem}
\end{figure}

A multilayer \lsmo--\sto~sample was characterized with scanning transmission electron microscopy (STEM), to study the interface atomic structure. The STEM data presented was measured with a FEI Titan microscope. The sample is comprised of several layers, from substrate to surface: 4 nm \lsmo, 2 nm \sto, 16 nm \lsmo, 2 nm \sto~and 8 nm \lsmo. 

Figure~\ref{stem}a presents a low magnification HAADF image of the different layers in the sample. In HAADF microscopy, the observed intensity of a column of atoms scales with the atomic weight of the elements within the columns. Therefore, the \lsmo~is brighter in the image than the \sto. Figure~\ref{stem}b presents a higher magnification image obtained from the layers close to the substrate. Both the $A$- and $B$-site columns are observed. The multilayer structure is grown coherently and following a quantitative analysis by statistical parameter estimation \cite{Aert2009}, it can be concluded that the interfaces are well defined with a chemical roughness of maximum 1 unit cell. 

The \lsmo--\sto~heterostructures have not only excellent structural properties, but they also have excellent functional properties. Thicker films of 10 nm show a Curie temperature of 350 K, a low-temperature saturation magnetization of 4 $\mu$$_\textrm{B}$/Mn, and a residual resistivity of 60 $\mu\Omega$cm. The heterostructures compare favourably to the samples presented in the literature \cite{Boschker2011}. The combination of high-quality films with excellent functional properties and the two-dimensional layer-by-layer growth allows for the interface engineering described in the remainder of this chapter. However, because atomic control of the crystal termination during growth is only possible up to a finite film thickness, reliable interface engineering can only be performed for thin \lsmo~heterostructures.

\section{Polar discontinuities at the interface}
\label{polar}
\begin{figure}[!htbp]
\centering
\includegraphics*[width=10cm]{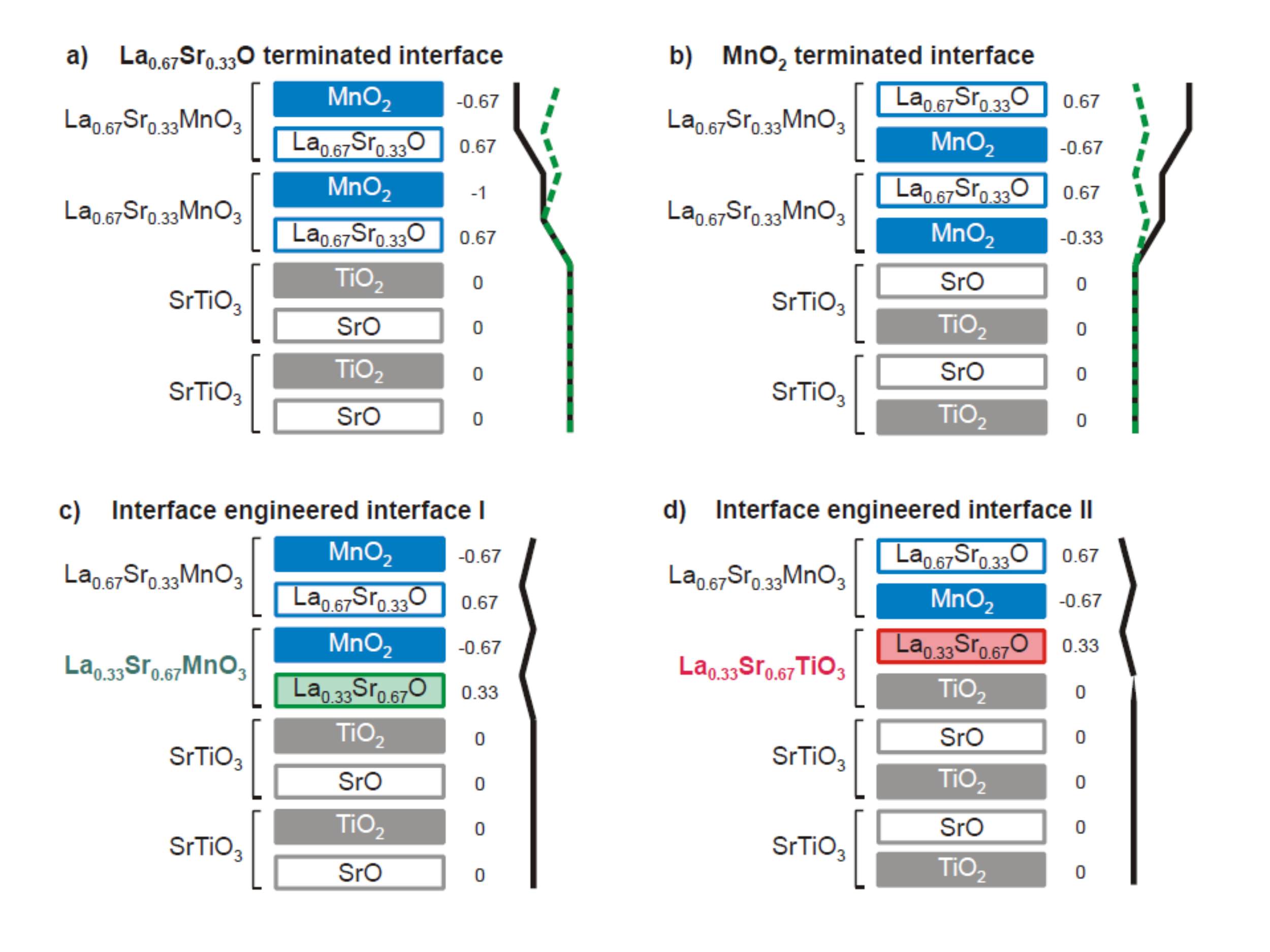}
\caption{The \lsmo--\sto~interface configurations. a) La$_{0.67}$Sr$_{0.33}$O terminated interface. b) MnO$_2$ terminated interface. c) Interface engineered TiO$_2$/La$_{0.33}$Sr$_{0.67}$O/MnO$_2$ interface I. d) Interface engineered TiO$_2$/La$_{0.33}$Sr$_{0.67}$O/MnO$_2$ interface II. The electrostatic potential due to the polar discontinuity is indicated with a black line (before reconstruction) and a green dotted line (after reconstruction). The numbers indicate the net charge at each layer, after reconstruction. Figure reprinted from \cite{Boschker2012}. } 
\label{lsmosto1}
\end{figure}

\lsmo--\sto~interfaces are polar and therefore ionic or electronic reconstructions are to be expected. Two different atomic stacking sequences are possible for the \lsmo--\sto~interfaces, due to the possible $A$O or $B$O$_2$ termination of each material. One will be referred to as the La$_{0.67}$Sr$_{0.33}$O terminated interface and the other as the MnO$_2$ terminated interface. The La$_{0.67}$Sr$_{0.33}$O terminated interface is shown in Fig.~\ref{lsmosto1}a. At this interface the atomic stacking sequence is SrO-TiO$_2$-La$_{0.67}$Sr$_{0.33}$O-MnO$_2$. The MnO$_2$ terminated interface is shown in Fig.~\ref{lsmosto1}b. Here the atomic stacking sequence is TiO$_2$-SrO-MnO$_2$-La$_{0.67}$Sr$_{0.33}$O. Both interface configurations are polar, and without reconstruction would result in a diverging electrostatic potential, as indicated in the figures. As \lsmo~is conducting, the mobile charges can screen the diverging potential. This screening, however, is only partial due to the fact that the charges are confined to the Mn sublattice. In Fig.~\ref{lsmosto1}a and \ref{lsmosto1}b the electrostatic potential after the screening (effectively an electronic reconstruction of the polar discontinuity) is shown as well. Here it is assumed that the screening occurs completely in the first MnO$_2$ layer, which is consistent with the Thomas Fermi length of \lsmo~of $\approx$~0.31 nm \cite{Hikita2009}. The diverging potential is eliminated, but a band offset, especially in the case of the La$_{0.67}$Sr$_{0.33}$O terminated interface, is present and will result in intermixing of the cations \cite{Nakagawa2006}. Note that the Ti$^{4+}$ valence state is stable at all these interfaces, as the Ti t$_{2g}$ levels are 1 eV higher in energy compared with the Mn e$_g$ levels \cite{Kumigashira2006apl}. 

In order to remove the polar discontinuity, an sub-unit-cell atomic layer with a net charge of 1/3 $e$-/uc has to be inserted at the interface. This is shown in Fig.~\ref{lsmosto1}c and \ref{lsmosto1}d, in which two methods are shown for interface engineering by inserting a single atomic layer of La$_{0.33}$Sr$_{0.67}$O between the two materials through the deposition of either a La$_{0.33}$Sr$_{0.67}$MnO$_3$ layer (I) or a La$_{0.33}$Sr$_{0.67}$TiO$_3$ layer (II). This gives an atomic stacking sequence at the interface of SrO-TiO$_2$-La$_{0.33}$Sr$_{0.67}$O-MnO$_2$-La$_{0.67}$-Sr$_{0.33}$O. In this case, the diverging potential is absent and no driving force for reconstruction exists. A band offset is present at this interface as well, so some intermixing can still be expected. 

Several heterostructures with and without interface engineering (denoted IE and non-IE, respectively) were grown. The structures consist of a TiO$_2$ terminated STO substrate, a variable thickness \lsmo~layer and a \sto~cap layer with a thickness of five unit cells. In order to realize the interface engineered heterostructure, first a single unit cell layer of La$_{0.33}$Sr$_{0.67}$MnO$_3$ was grown. This was followed with the growth of $n$-1 layers of \lsmo, then a single unit cell layer of La$_{0.33}$Sr$_{0.67}$TiO$_3$ and finally, four unit cell layers of \sto. This recipe resulted in heterostructures with $n$ layers of Mn ions, which can therefore be compared with the $n$ unit cell layer \lsmo--\sto~structure without the interface engineering. Next to the single layer samples, also two superlattice samples (IE and non-IE) with five repetitions of (\lsmo)$_8$(\sto)$_5$ were grown. 

\begin{figure}[!htbp]
\centering
\includegraphics*[width=10cm]{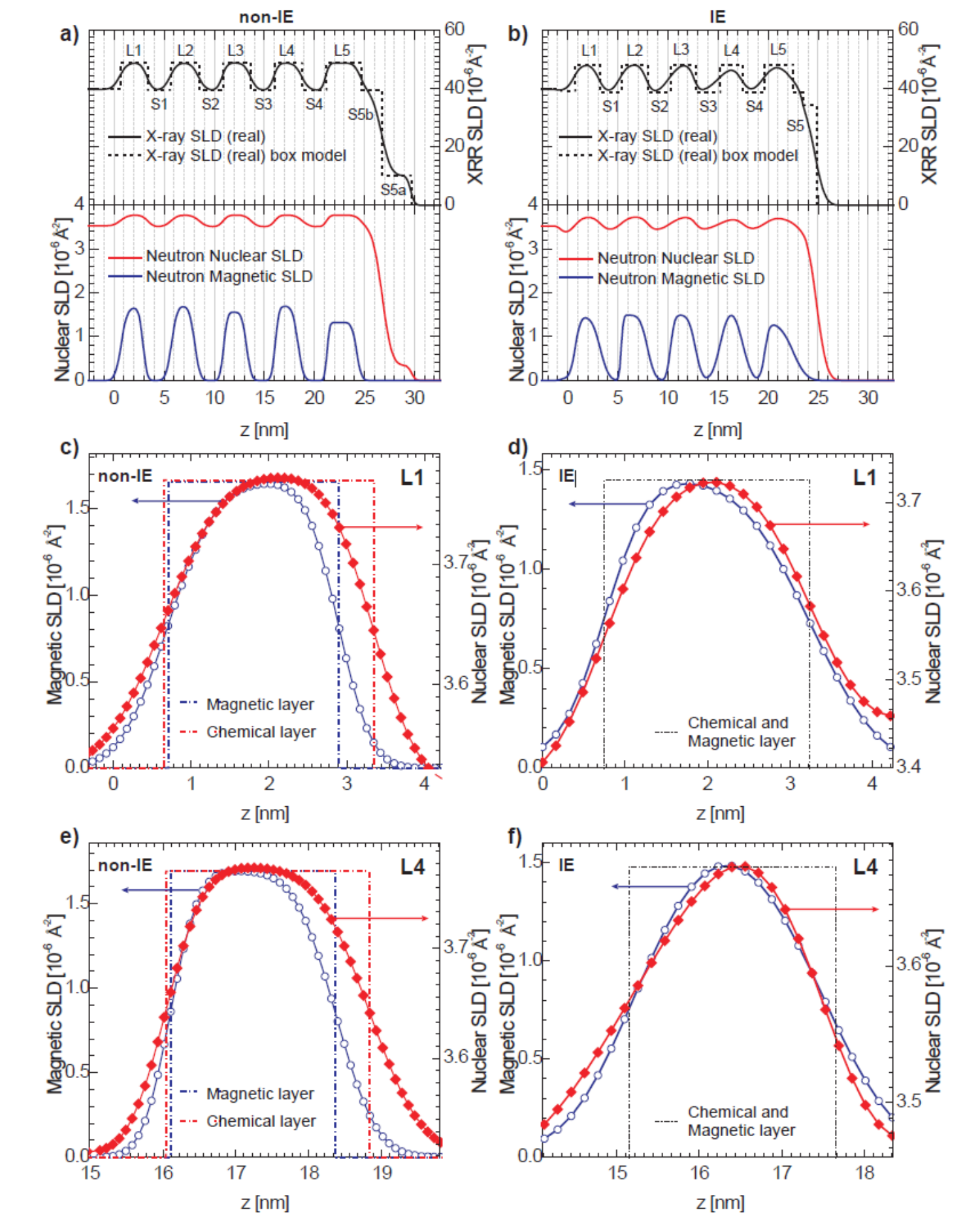}
\caption{ The magnetic and chemical scattering length density (SLD) profiles at various film thickness $z$ obtained from fitting neutron and x-ray reflectivity data for \lsmo--\sto~heterostructures a) without and b) with interface engineering. The dashed lines represent the SLD profiles with sharp interfaces (a so-calles box model). c-f) Overlay of the chemical and magnetic profile without c,e) and with d,f) interface engineering for respectively the first (L1) and fourth (L4) \lsmo~layer. Figure reprinted from \cite{Huijben2015}.} 
\label{pnr}
\end{figure}

The samples were analyzed extensively by STEM and electron energy loss spectroscopy (EELS) \cite{Boschker2012, Huijben2015}. It was found that in the non-IE samples differences between the top and bottom interface exist, due to the difference in stacking sequence at the interface. At the bottom interface (La$_{0.67}$Sr$_{0.33}$O terminated interface) a relatively large diffusion of La ions into the \sto~was found which is probably the result of a reconstruction of the polar discontinuity. In contrast, the IE samples show similar top and bottom interfaces, in agreement with the design of the heterostructures. 

The depth profiles of the magnetization in the superlattice samples was measured with polarized neutron reflectometry (PNR) \cite{Huijben2015}. The main result of the analysis is shown in Fig.~\ref{pnr}. These profiles show that the topmost \sto~and \lsmo~layers exhibited different structural properties, most likely due to relaxation at the surface. However, all other four buried \lsmo--\sto~bilayers showed to be structurally equal as demonstrated by showing the first (L1) and fourth (L4) \lsmo~layer in respectively Figs.~\ref{pnr}c,e and \ref{pnr}d,f. The chemical roughnesses in the IE heterostructure are typically 1.5 times larger than that in the non-IE heterostructure. This can to some degree be attributed to the more gradual La distribution in the design of the IE heterostructure. The magnetic roughnesses, in contrast, are similar in both heterostructures.

Most importantly, for the non-interface-engineered (non-IE) heterostructure a clear 5.5 \AA~$\pm$ 1.0 \AA~difference between the magnetic and chemical thickness of the \lsmo~layer was observed (Figs.~\ref{pnr}c and \ref{pnr}e). The optimal model indicates that only the top part of each \lsmo~layer was strongly magnetically reduced. Because the La profile in the non-IE structures is shifted with respect to the Mn profile (STEM-EELS measurements), we also tested an alternative model with equally sized dead-layers at both interfaces. The model also yields a total 5 \AA~magnetically dead region (2.5 \AA~each at top and bottom interfaces), similar to the optimal model, albeit with a slightly increased fitting error. Note that the difference between magnetic and chemical profiles at the bottom interface is largely due to La diffusion, as the first Mn layer is still magnetic. On the other hand, at the top interface, a reduction of the magnetization is found in a layer where Mn exists, therefore a true magnetic dead-layer \lsmo~only resides at the top interface. This is in good agreement with a previous observation that a surpluss of La does not deteriorate the magnetism \cite{Kavich2007}. Overall, in the non-interface engineered heterostructure, the magnetic thickness is $\sim$5 \AA~less than the chemical thickness, and there is at least a magnetically reduced region at the top, and possibly at the bottom, the combination of which is 5.5 \AA~$\pm$ 1.0 \AA~thick.

In contrast, the interface-engineered (IE) heterostructure demonstrated an equal magnetic and chemical thickness of the \lsmo~layer (Figs. \ref{pnr}d and \ref{pnr}f). The integrated magnetization at 120 K over the eight unit-cell-thick \lsmo~layer, extracted from the PNR data, is $\sim$3.3 $\mu$$_\textrm{B}$/Mn for the heterostructure with interface engineering, while a heterostructure without interface engineering exhibited a magnetization of only $\sim$2.8 $\mu$$_\textrm{B}$/Mn. The absolute magnetization values extracted from PNR are confirmed by VSM measurements and the observed magnetization enhancement of 18 \% is in very good agreement with the observed increase in magnetization of a single \lsmo--\sto~bilayer when interface engineering was applied.

\begin{figure}[!htbp]
\centering
\includegraphics*[width=8cm]{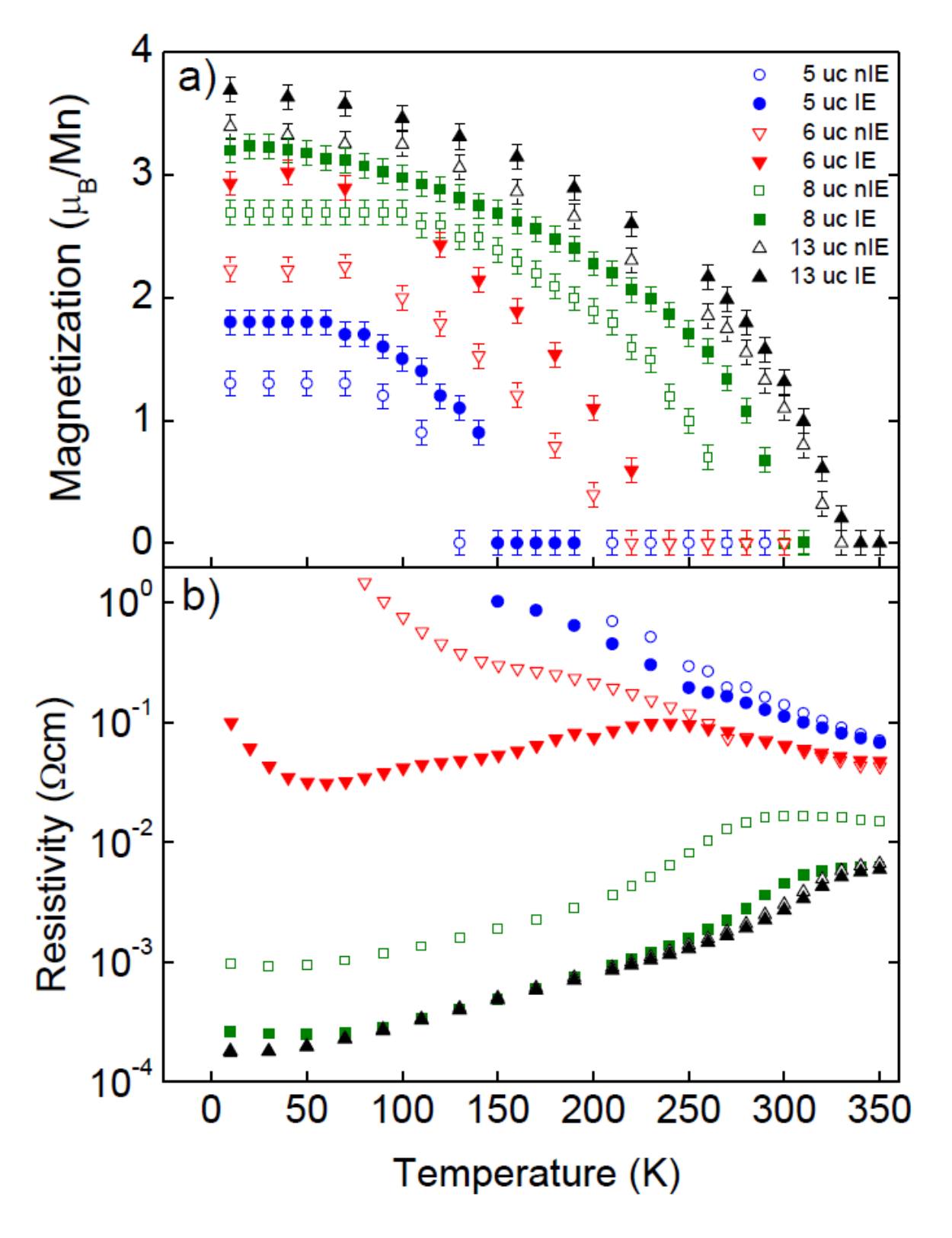}
\caption{ Temperature dependent a) magnetization and b) resisitivity measurements for \lsmo--\sto~heterostructures of various $n$ thicknesses in unit cells (uc). Results of samples with (IE) and without interface engineering (nIE) are shown by respectively closed and open symbols. Figure reprinted from \cite{Boschker2012}.} 
\label{ietransport}
\end{figure}

Figure \ref{ietransport} presents temperature-dependent magnetization and resistivity measurements of the bilayer samples. The magnetic field was applied in-plane along the [100]$_\textrm{c}$ \sto~crystal direction during the magnetization measurements. Figure \ref{ietransport}a shows the temperature dependence of the saturation magnetization
for the $n$ = 5, 6, 8 and 13 uc samples. A clear difference between the IE and non-IE samples is observed. The IE samples have significantly higher saturation magnetization and Curie temperature ($T_\textrm{C}$). Figure \ref{ietransport}b presents the temperature dependence of the resistivity of the samples. For all thicknesses, the interface engineering results in a lower
resistivity. The $n$=5 samples and the non-IE $n$=6 sample are insulating at low temperature. The $n$=6 IE sample has an upturn in the resistivity at low temperature but conductivity was observed down to the lowest temperature (10 K) in the measurement. An enhancement of the magnetization and electrical conductivity can still be observed for thicker \lsmo~layers in the 13 unit-cell-thick samples, however, the effect is smaller as the enhanced interface properties provide a smaller contribution to the properties of the complete \lsmo~layer when thicker layers are studied. The obtained properties of the non interface engineered 13 uc sample are in good agreement with previously obtained results in ultrathin \lsmo~films \cite{Huijben2008}.

These results provide strong evidence of the elimination of a magnetic dead-layer at the \lsmo--\sto~interface when applying interface engineering by incorporating of a single La$_{0.33}$Sr$_{0.67}$O layer, which leads to embedding of the interfacial MnO$_2$ atomic layer in between two (La,Sr)O layers. Furthermore, both the interface asymmetry of the magnetically reduced layer in the non-IE heterostructure and the absence of the dead-layer in the IE heterostructure confirm the reconstructions induced by the polar discontinuity at the interface as the origin of the former magnetic dead-layer at \lsmo--\sto~interfaces. However, even in the IE samples, the electrical dead-layer is still present and the question remains whether we can engineer even better interfaces. 

\section{Order and disorder}
\label{order}

\lsmo~is a solid solution of \lmo~and \smo. This implies \lsmo~is an intrinsically disordered material, where the $A$-site ion on a certain lattice site can be occupied with either a La or a Sr ion. It is an interesting question whether the disorder affects the properties, e.g. electrical transport. Several studies have presented results on the properties of (\lmo)$_{2n}$/(\smo)$_{n}$ superlattices, which have an identical La/Sr ratio with respect to \lsmo~\cite{Yamada2006, Adamo2008, Bhattacharya2008}. A decrease of the residual resistivity, with respect to the solid solution, was observed for the $n=1$ superlattice \cite{Bhattacharya2008}. 

Here, the effect of the disorder on the properties at the interface between \lsmo~and \sto~is studied. It is proposed that the \sto--\smo~interface has no intrinsic disorder, as the $A$-site ions of both materials are equal. To obtain \lsmo's functional properties at the interface, charge has to be supplied to the \smo~layer. This is accomplished by chemically doping the \sto~away from the interface. This results in modulation doping \cite{Dingle1978}, in which charge is transferred from the doped \sto~layer to the \smo~layer. The amount of charge which can be transferred is limited by the potential drop due to the electric dipole between the transferred electrons and the positively charged dopants. Eventually, the potential drop increases the energy of the \smo~conduction band to the energy of the \sto~conduction band and then the charge transfer stops. The maximum charge transfer is then given by:
\begin{equation}
x=\frac{\Delta \epsilon_0 K}{d},
\label{moddopct}
\end{equation}
in which $x$ is the maximum charge transfer, $\Delta$ is the energy difference between the \sto~and \smo~conduction bands, $\epsilon_0$ is the permittivity of free space, $K$ is the dielectric constant and $d$ is the distance over which the charge transfer occurs. Therefore, the use of \sto~has a great advantage; its huge dielectric constant screens the potential drop. It is possible to dope two $e$-/uc over a distance of two nm, which is an order of magnitude larger as the charge transfer by modulation doping in semiconductor structures. And at low temperatures, significant charge transfer can be present without significant band bending, as the dielectric constant of \sto~increases massively with decreasing temperature. 

\begin{figure}[!htbp]
\centering
\includegraphics*[width=8cm]{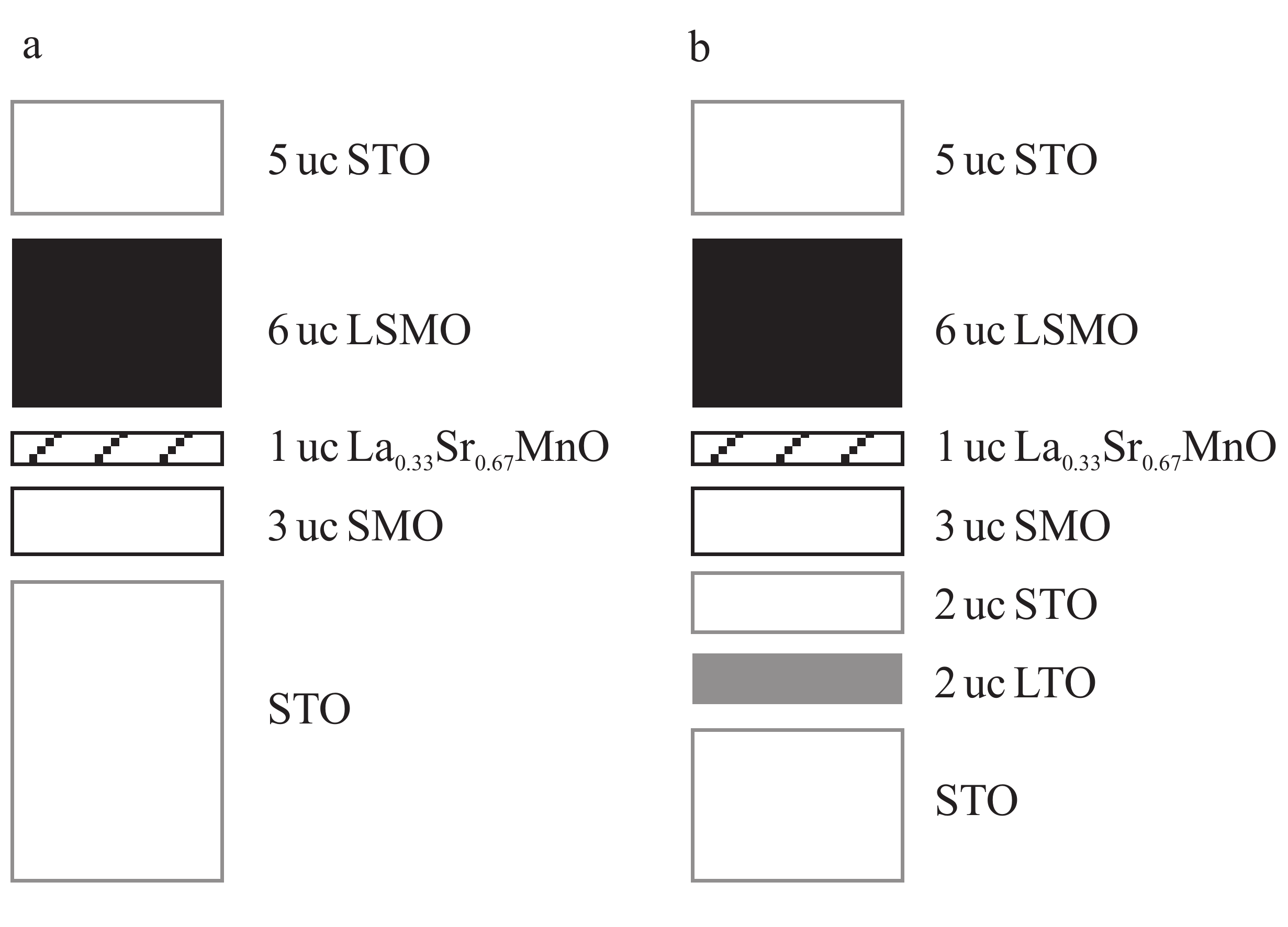}
\caption{A schematic of the sample structure. a) reference sample A. b) modulation doped sample B.}
\label{moddopsam}
\end{figure}

We tested the modulation doping scheme by growing two samples with ten unit-cell-thick manganite layers. A schematic image of the two samples is shown in Fig.~\ref{moddopsam}. Sample A is a reference sample, where a \smo~layer is inserted between the \lsmo~and the \sto~to obtain the \sto--\smo~interface. The thickness of the \smo~layer is three unit cells, which corresponds to the amount of electrically dead layers at the \lsmo--\sto~interface. Next, a one unit-cell-thick layer of La$_{0.33}$Sr$_{0.67}$MnO$_{3}$~is present between the \smo~and the \lsmo, as this results in higher quality growth of the \lsmo. The \lsmo~layer thickness is six uc, so the sample can be compared with an \lsmo~sample with a thickness of ten unit cells. For this sample no dopants are present in the \sto, so the \lsmo~layer is expected to be overdoped. Sample B is similar to sample A, but here a two unit-cell-thick layer of LaTiO$_3$~is inserted in the \sto~two unit cells away from the interface. The LaTiO$_3$ layer dopes two $e$-/uc to the structure, which is precisely the amount required in the \smo~layer to obtain \lsmo's Mn$^{3.33+}$ valence state. Note that only the bottom interface has been engineered to remove the disorder, while the top interface is a disordered \lsmo--\sto~interface.

\begin{figure}[!htbp]
\centering
\includegraphics*[width=10cm]{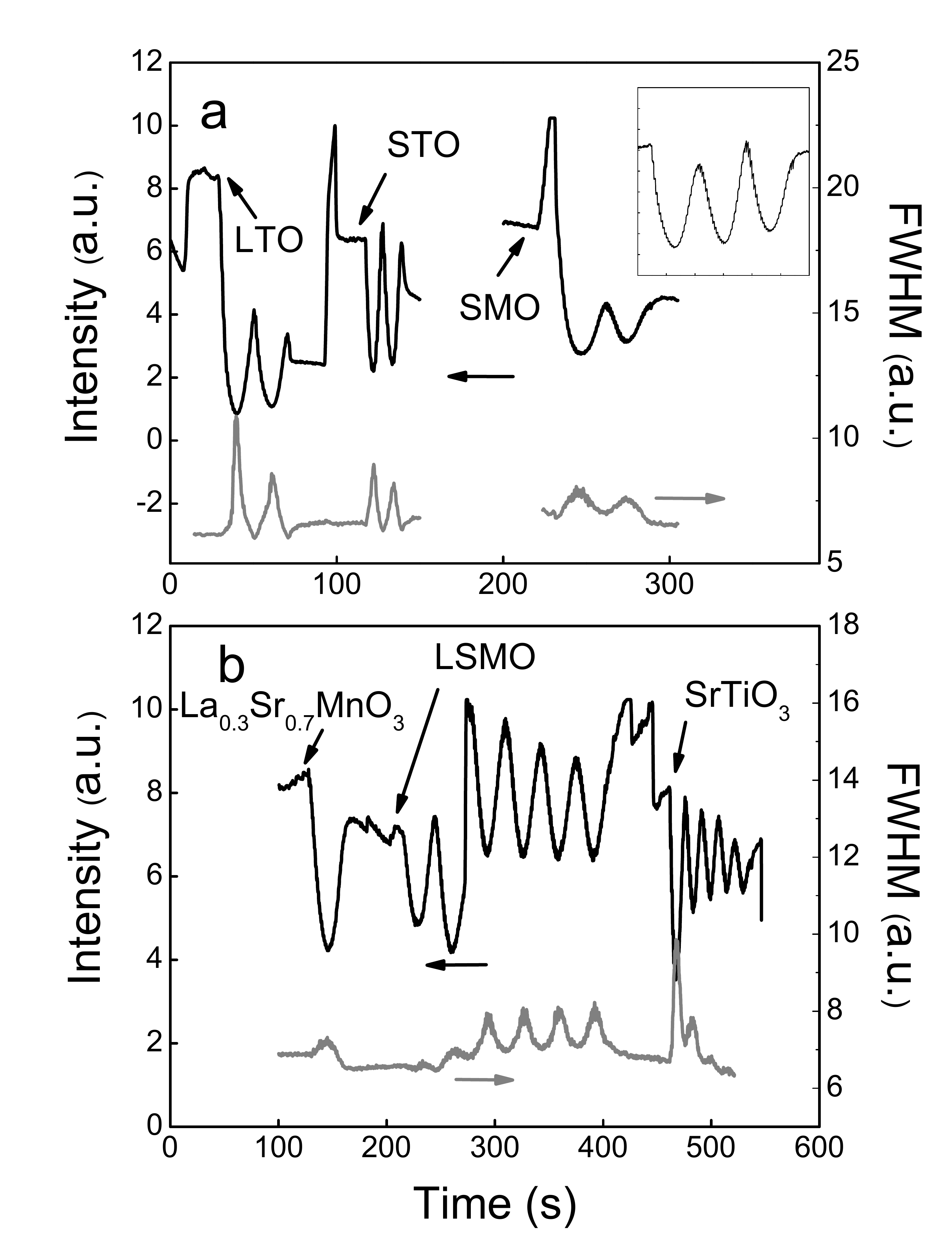}
\caption{RHEED specular spot intensity oscillations and FWHM during the growth of sample B. The starting point of each layer is indicated. The inset in a) shows the RHEED specular spot intensity observed during the growth of \smo~directly on the \sto~substrate, sample A.  }
\label{moddoprheed1}
\end{figure}

The intensity of the specular spot, as monitored with RHEED during the growth of sample B, is shown in Fig.~\ref{moddoprheed1}. Clear oscillations are observed during the growth of LaTiO$_3$ and \sto. The initial growth of \smo~on the LaTiO$_3$--\sto~stack started with an increase of the RHEED intensity to a maximum, followed by a prolonged oscillation. This second maximum occurred at the expected time for the deposition of 1.5 unit cell layers. The period of the second oscillation corresponds to the time required to deposit a single unit cell layer. As the \smo~deposition was stopped at the third RHEED intensity maximum, only 2.5 unit cells of \smo~instead of three were grown. This initial growth of \smo~was not observed for the \smo~growth directly on the \sto~substrate. The inset of Fig.~\ref{moddoprheed1}a shows the RHEED specular spot intensity observed during the \smo~growth of sample A. Here three clear oscillations with equal periods are observed. After the growth of the \smo, the growth of the La$_{0.33}$Sr$_{0.67}$MnO$_{3}$, \lsmo~and \sto~layers showed RHEED oscillations. Therefore, it is concluded that the entire heterostructure was grown in 2D layer-by-layer fashion.

\begin{figure}[!htbp]
\centering
\includegraphics*[width=10cm]{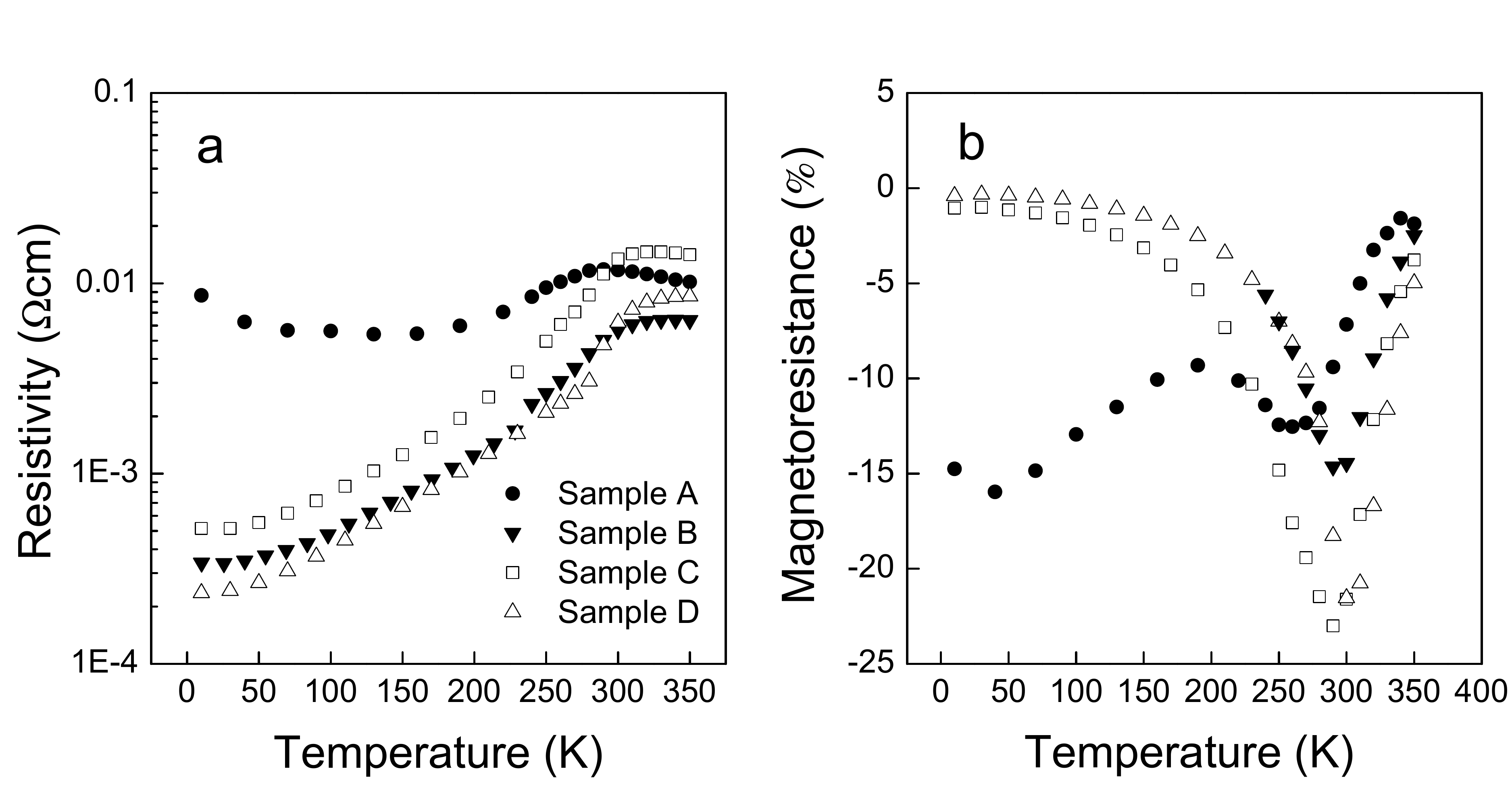}
\caption{Temperature dependent resistivity measurements of four ten-unit-cell-thick \lsmo~samples. Sample A contains a \smo~layer, but has no modulation doping. Sample B contains the \smo~layer and has modulation doping. Sample C is a standard \lsmo~sample, capped with \sto~and sample D is similar to sample C, but with engineered interfaces to remove the polar discontinuity.}
\label{moddopres}
\end{figure}

Electrical transport measurements of the $n=10$ samples are presented in Fig.~\ref{moddopres}. For comparison, also two other samples are shown. Sample C is a ten unit cell layer \lsmo~sample capped with five unit cell layers of \sto. Sample D is similar as sample C, but with engineered interfaces to remove the polar discontinuities, as discussed in section~\ref{polar}. The resisitivity of sample B is similar to the values observed for samples C and D, indicating that sample B has the properties of a ten unit cell layer \lsmo~sample. Sample A, in contrast, has a higher resistivity, lower $T_\textrm{C}$ and negative magnetoresistance present at all temperatures. The transport behaviour of sample A is actually similar to that of seven unit-cell-thick \lsmo~samples, consistent with the presence of seven unit cells of \lsmo~in the sample structure. This result indicates that charge is transferred from the LaTiO$_3$ layer to the \smo~layer and that the charge transfer significantly changes the conductivity.

\begin{figure}[!htbp]
\centering
\includegraphics*[width=8cm]{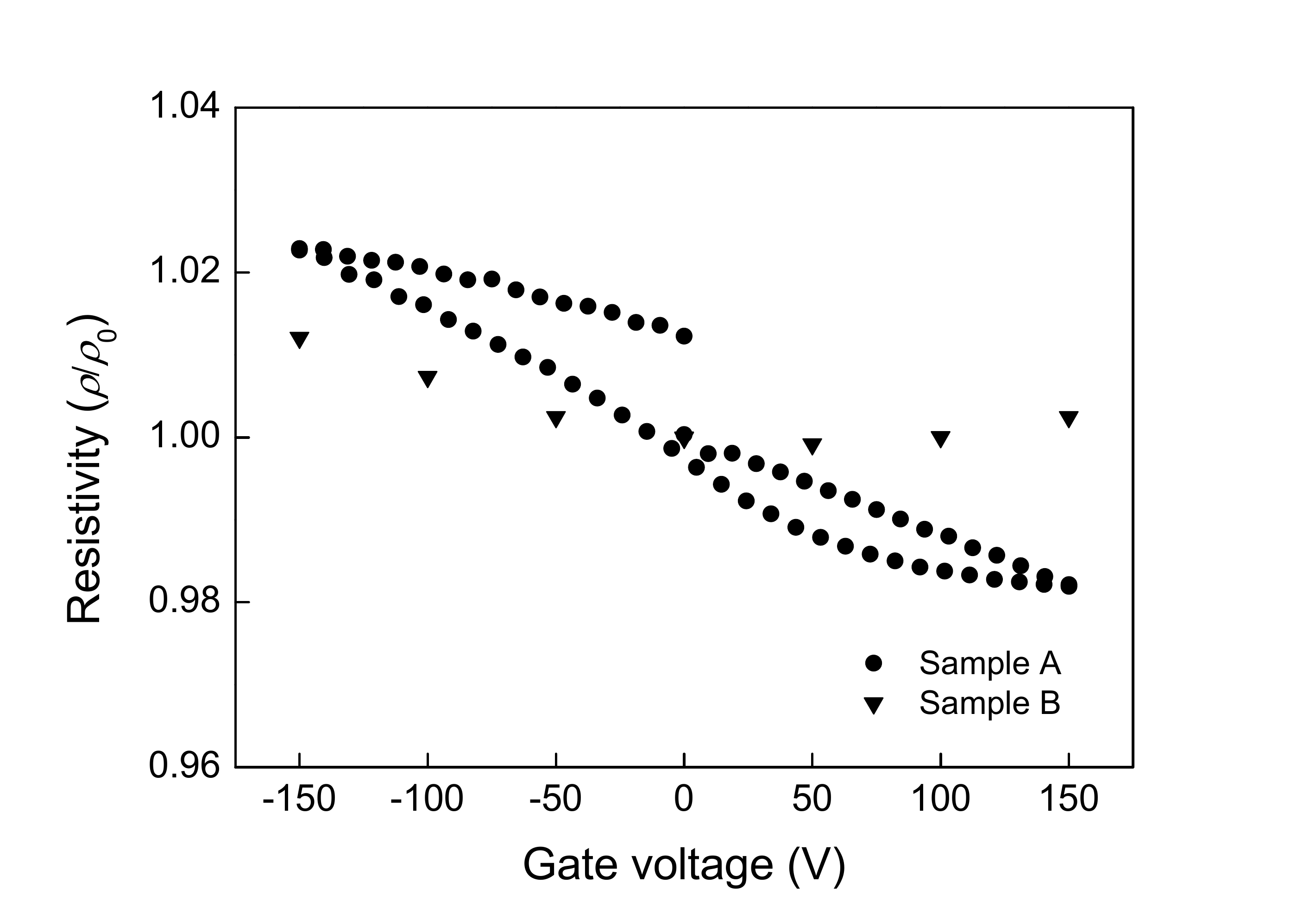}
\caption{The gate voltage dependence of the resistivity. }
\label{moddopgate}
\end{figure}

In order to verify that the optimum amount of charge is present in the \smo~layer, electrical field effect measurements were performed. The charge carrier density of the sample can be controlled with the gate voltage applied to the back of the sample. Figure~\ref{moddopgate} presents the gate voltage dependence of the resistivity, measured at 10 K. Sample A shows some hysteretic behaviour, probably due to charge impurities in the \sto~substrate, but the general dependence is monotonic with gate voltage. A decrease of the resistivity is observed for positive gate voltage (electron accummulation in the \smo). This corresponds well with the expected overdoping of sample A. For sample B, a minimum in the resistivity is observed at a gate voltage of 50 V. This indicates that the optimum charge carrier density for the conductivity of the \lsmo~is very close to the doping achieved with the modulation doping. With a \sto~dielectric constant of 20000, at most 0.015 $e$-/uc charge carriers are accummulated by the electric field. From a practical point of view, the control over the amount of dopants is 5\% (approximately 1 laser pulse during deposition) of a unit cell layer of LaTiO$_3$, which corresponds to 0.05 $e$-/uc. Therefore, it is concluded that sample B is as good as can presently be realized and little room for improvement exists in optimizing the inserted amount of dopants. The experiments indicate that modulation doping with a significant charge transfer was achieved. However, \lsmo's conductivity at the interface was not much improved, indicating that disorder is not the key parameter governing the properties at the interface.

\section{Octahedra rotations at the interface}
\label{liao}
In the preceeding sections we presented improvements of the \lsmo~interface properties by removing the polar discontinuities and the intrinsic disorder at the interfaces. However, the electrical dead-layer still remains. The likely origin of the dead-layer lies in the structural reconstructions at the interface \cite{Pruneda2007, He2010}. In this section we study the structure of the oxygen octahedra at the interface in detail. 

The changes in the unit-cell symmetry resulting from different octahedral rotations have been systematized by Glazer \cite{Glazer1972} and later expanded by Woodward \cite{Woodwarda, Woodwardb}. Octahedral rotations in the perovskite-type unit cell can be described as a combination of rotations about three symmetry axes of the pseudocubic unit cell. The relative magnitudes of the tilts are denoted by letters a, b, and c, $e$.$g$., aab means equal rotations around the $[100]$ and $[010]$ axes and a different tilt around the $[001]$ axis. Two adjacent octahedra around one of the $<$100$>$ axes can rotate either in-phase or out-of-phase, which is indicated by the + or -- sign, respectively. No rotation is indicated by the 0 sign.

\sto~is a cubic crystal (above 105 K) with Glazer tilt system a$^0$a$^0$a$^0$. \lsmo~is rhombohedral with tilt system a$^-$a$^-$a$^-$. \lsmo~films grown on \sto, however, have tilt system a$^+$a$^-$a$^0$ in order to accommodate the tensile strain applied to the films \cite{Vailionis2011}. Clearly a tilt system mismatch exists at the interface and because the oxygen octahedra network has to be continuous across the interface, structural distortions of the octahedra have to be present. These distortions have been predicted to occur on the \lsmo~side of the interface \cite{He2010}.

\begin{figure}[!htbp]
\centering
\includegraphics*[width=10cm]{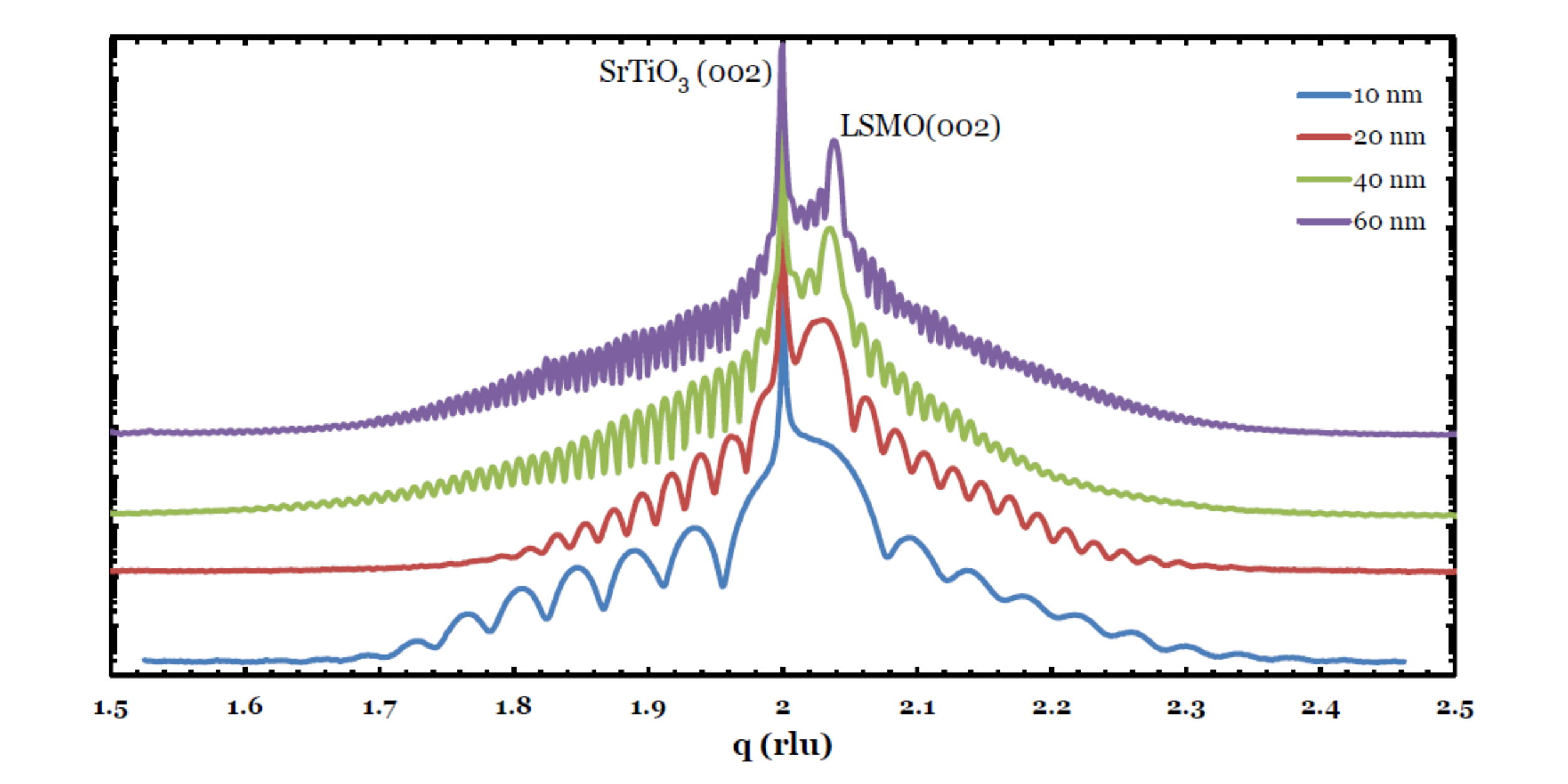}
\caption{Synchrotron XRD data for \lsmo~films with different thicknesses. }
\label{fringes}
\end{figure}

\begin{figure}[!htbp]
\centering
\includegraphics*[width=10cm]{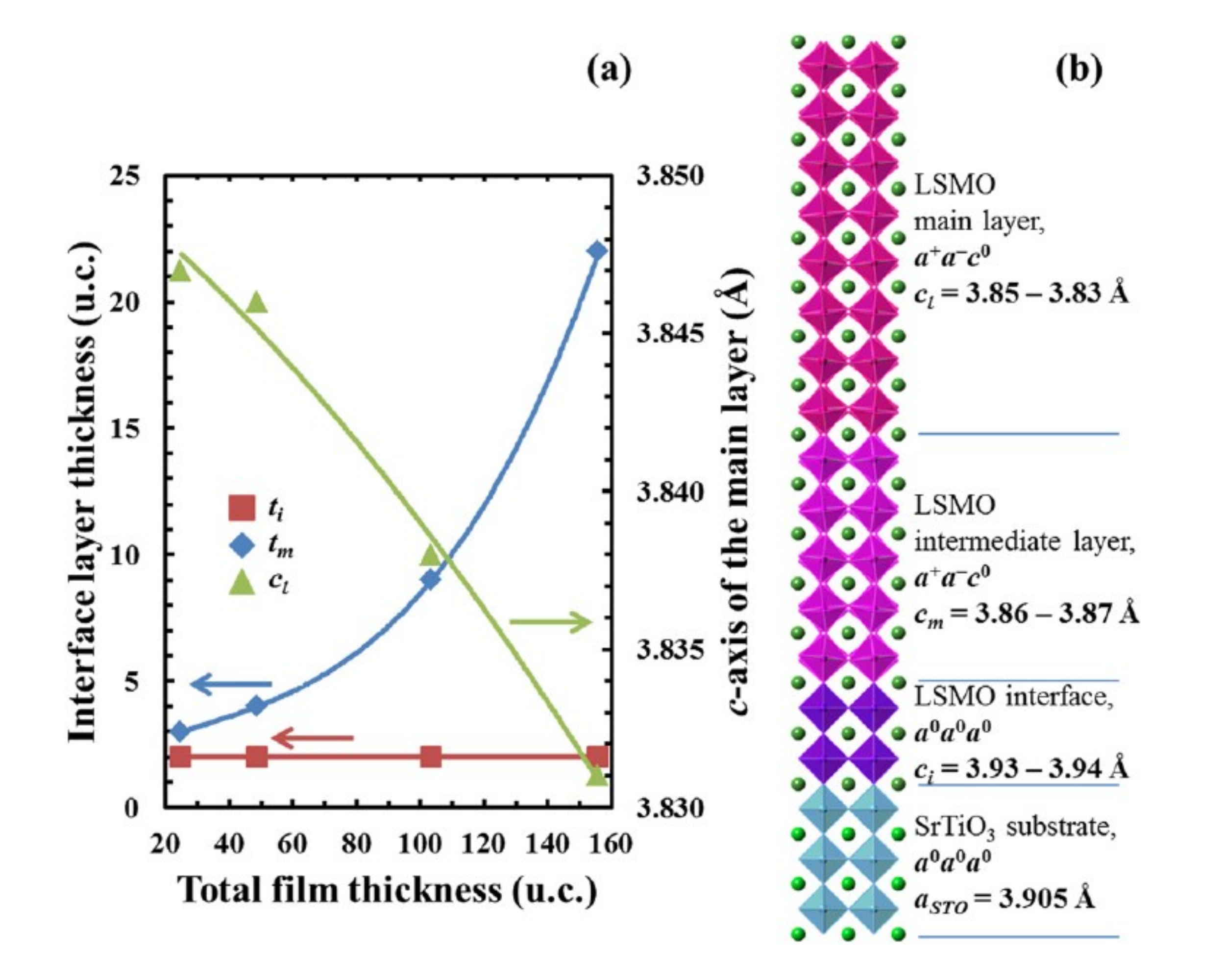}
\caption{ Structural evolution of the \lsmo~thin film along the out-of-plane direction. a) Variation of the interface and intermediate layer thickness and $c$-axis lattice parameter of the main layer as a function of the total film thickness. b) Schematic diagram of the \lsmo~layer structural evolution near and away from the \lsmo--\sto~interface. Figure reprinted from \cite{Vailionis2014}. }
\label{xrdresult}
\end{figure}

Four \lsmo~samples with different thicknesses (10, 20, 40 and 60 nm) were studied with synchrotron x-ray diffraction (XRD). Figure~\ref{fringes} shows out-of-plane momentum scans around the (002) Bragg reflection. Next to the Bragg peaks, fringes are present due to the finite thickness of the samples. Due to the low surface and interface roughnesses in the samples a large number of fringes is seen and quantitative analysis is possible. The data contains two features that are inconsistent with the \lsmo~having a uniform thickness throughout the sample: the fringe intensity at small momentum transfer (with respect to the Bragg peak) is larger than that at high momentum transfer and the fringe intensities exhibit a modulation with a larger period. Dynamical XRD calculations can only reproduce these features when several layers with different out-of-plane lattice parameters are included in the model of the sample structure \cite{Vailionis2014}. Three distinct layers are present in the samples: an interface layer, an intermediate layer and the main layer, see Fig.~\ref{xrdresult}.

Here we focus on the interface layer. Surprisingly we find that the interface layer is two unit cells thick with an out-of-plane lattice parameter of 3.93-3.94 \AA, independent of the total film thickness. This implies that the first atomic layers of the films are under compressive strain as opposed to the tensile strain in the rest of the films. The result can easily be understood by considering the interface mismatch in the oxygen octahedra rotation patterns between \lsmo~and \sto. The Mn-O bond length in \lsmo~is 1.957 \AA, so with the a$^0$a$^0$a$^0$ tilt pattern of \sto, the \lsmo~would be cubic with a lattice constant of of 3.914 \AA. This is larger than the lattice constant of \sto~and the \lsmo~interfacial layers are under compressive strain. The strain distorts the Mn-O bond length and, with a Poisson ratio of 0.4, the out-of-plane lattice parameter would be 3.93 \AA, consistent with the experiment. In conclusion, the first two unit cells of \lsmo~at the \lsmo--\sto~interface are severely distorted due to structural coupling to the \sto. This interfacial layer has an a$^0$a$^0$a$^0$ tilt pattern and the oxygen octahedra are elongated along the out-of-plane direction. We note that structural distortions are also found in photoemission experiments at this interface \cite{Gray2010, Gray2013}. The elongation of the octahedra at the interface breaks the orbital symmetry and thereby the double exchange interactions are weakened at the interfaces. 

\begin{figure}[!htbp]
\centering
\includegraphics*[width=10cm]{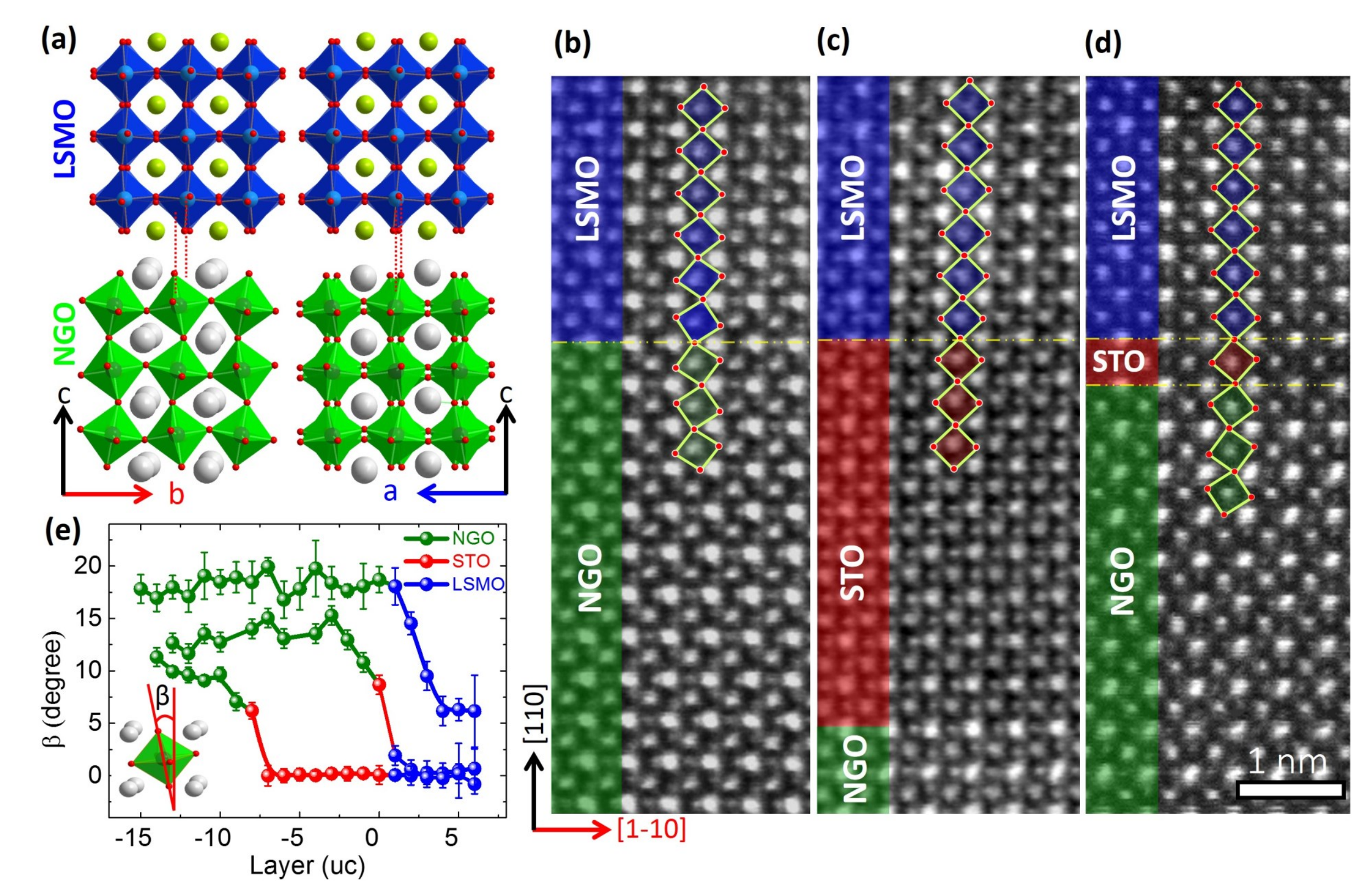}
\caption{ Oxygen octahedral coupling at interfaces in manganite heterostructures. (a) Schematic models of atomic ordering in \lsmo~and \ngo~crystal structures. (b-d) Inversed annular bright-field STEM images of \lsmo--\ngo~(left) and \lsmo--\sto--\ngo~(right) heterostructures. The oxygen atoms are clearly visible, and the connectivity of oxygen octahedra across the interfaces is indicated. The MnO$_6$ octahedra show a clear in-phase rotation following the \ngo. The \lsmo~films are 6 uc thick while the \sto~buffer layer has a thickness of 9 or 1 uc. (e) Layer-position dependent octahedron tilt angle ($\beta$) in \lsmo--\ngo~heterostructures with and without a \sto~buffer layer. Figure reprinted from \cite{Liao2016}. }
\label{STEMNGO}
\end{figure}

The situation is different in \lsmo--\ngo~heterostructures. \ngo~is orthorhombic with the a$^-$a$^-$c$^+$ tilt pattern, whereas \lsmo~grown on \ngo~(110) has the a$^+$b$^-$c$^-$ tilt pattern \cite{Vailionis2011}. These structural characteristics give rise to an in-phase vs. out-of-phase rotation type mismatch occurring along the [001]-axis that is absent along the [1$\overline{1}$0]-axis where both materials' rotations are out-of-phase. The magnitude of the bond angle $\theta$ also has a certain degree of mismatch: $\sim$154$^\circ$ in \ngo~vs. 166.3$^\circ$ in \lsmo. As a result, both the anisotropic rotation type mismatch and the large difference ($\sim$12$^\circ$) in bond angle will cause a strong discontinuity of the octahedra. In order to retain the connectivity of the corner shared oxygen octahedra across interface, the large octahedral tilt present in the \ngo~substrate will propagate into \lsmo~near interface region, similar to what occurs in \lsmo--\sto~interfaces as mentioned above. Figure~\ref{STEMNGO} presents STEM images of three \lsmo--\ngo~heterostructures together with schematics of the bulk structures of the materials. A direct visualization of the octahedra rotations is possible with Cs-corrected STEM via which the oxygen atoms can be imaged through the angular bright field technique. Figure~\ref{STEMNGO}b clearly shows that \ngo~induces large rotations of octahedra in the \lsmo~layers close to the interface. The octahedral tilt angle can be quantified by statistical parameter estimation theory \cite{Liao2016}, and is presented in Fig.~\ref{STEMNGO}e. The tilt angle continuously changes from the \ngo~substrate value to that of bulk \lsmo~(far from interface). Interestingly, the first two unit cell layers of the \lsmo~have almost the same tilt angle as the \ngo. The impact of the octahedral coupling decays rapidly away from the interface and disappears above four unit cell layers.
 
Because of the short impact length scale of oxygen octahedra coupling, the rotations of the \lsmo~can be significantly altered by inserting a non-tilted \sto~buffer layer (Fig.~\ref{STEMNGO}c). Within the \sto~layer, the oxygen octahedra rotations are also coupled to those of the \ngo, but the tilt angle relaxes quickly, i.e., the tilt of the TiO$_6$ octahedra disappears above two unit cell layers. Consequently, the \lsmo~connects to a non-tilted octahedra structure and therefore, within the STEM spatial resolution, there is no tilting of the MnO$_6$ octahedra (see Fig.\ref{STEMNGO}e), neither at the interface nor away from the interface. Due to steep decay of the tilts in the \sto, a one unit-cell-thick \sto~buffer layer suffices to significantly reduce the tilts in the \lsmo~(see Fig.\ref{STEMNGO}d,e). These results indicate that the local oxygen octahedra rotations at the substrate surface act as a controllable template for the structure of the epitaxial \lsmo~film.
 
\begin{figure}[!htbp]
\centering
\includegraphics*[width=10cm]{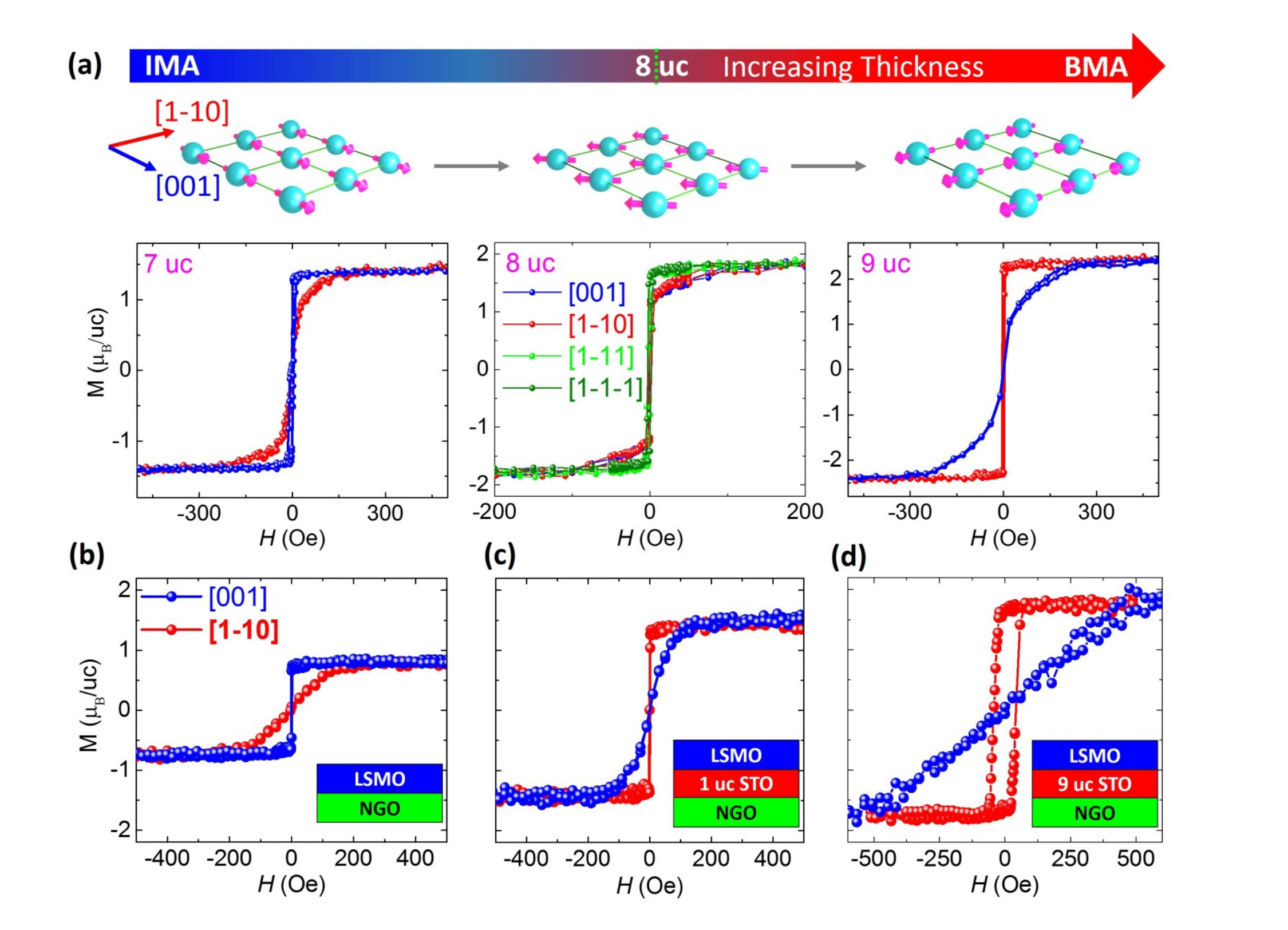}
\caption{ Atomic scale control of magnetic anisotropy. (a) The $M$($H$) curves at 100 K along the [001] and [1$\overline{1}$0] axis of \lsmo~films with thicknesses of 7, 8 and 9 uc on \ngo~substrates. The schematics at the top show the corresponding ground state of the Mn spin orientation. The $M$($H$) curves at 100 K along the [001] and [1$\overline{1}$0] axis of the 6 uc \lsmo~films on \ngo~substrates without (b) and with a one uc (c) and 9 uc (d) \sto~buffer layer. Figure reprinted from \cite{Liao2016}.}
\label{MagAn}
\end{figure}

The strong octahedral tilts in \lsmo~near interface in \lsmo--\ngo~heterostructures affects the magnetic properties when the films are thin enough for the interfacial structure to dominate. In strong contrast to the normally observed easy axis along the long [1$\overline{1}$0] axis \cite{Boschker2009}, the easy axis of ultrathin \lsmo~with $t$ $<$ eight unit cells is along the short [001] axis due to the interfacial magnetic anisotropy (IMA), as shown in Fig.~\ref{MagAn}a. The contribution of anisotropic strain to the magnetic anisotropy will increase with increasing thickness, hence thicker films ($t$ $>$ eight unit cells) exhibit bulk magnetic anisotropy (BMA) with the easy axis along the [1$\overline{1}$0] axis. At eight unit cells, the competition between the IMA and the BMA results in biaxial anisotropy with the easy axes along the [1$\overline{1}$1] and [1$\overline{11}$] axes. The engineering of octahedral tilting by introducing an ultrathin \sto~buffer layer (see Fig.~\ref{MagAn}b-d) also provides us with capabilities to switch the IMA to BMA in ultrathin LSMO films, e.g., a one unit-cell-thick \sto~buffered six unit-cell-thick \lsmo~film exhibits an easy axis along the [1$\overline{1}$0] axis (see Fig.~\ref{MagAn}c). Therefore, the short impact length scale of octahedra coupling becomes a unique tool that allows us to atomically engineer octahedra rotations at the interface and thus anisotropic properties at the atomic scale.

\begin{figure}[!htbp]
\centering
\includegraphics*[width=10cm]{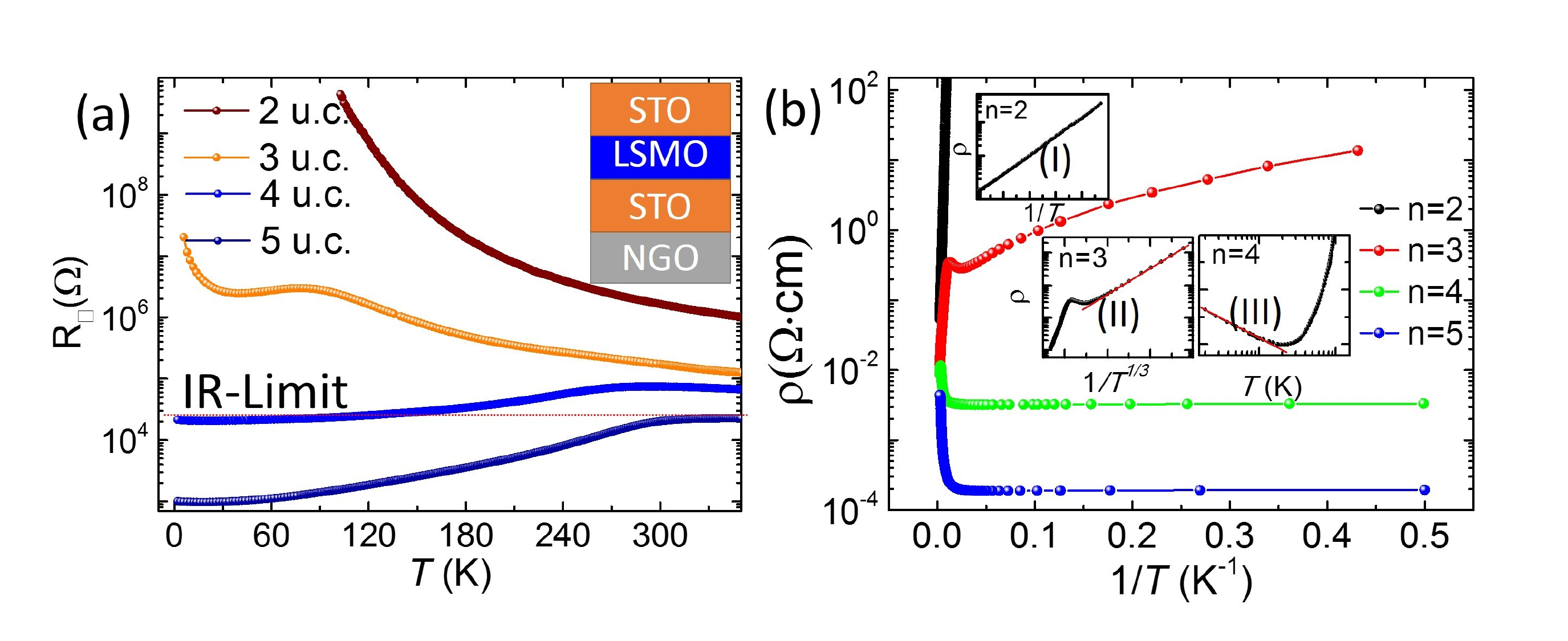}
\caption{ (a) Temperature dependence of the sheet resistance for different thicknesses of \lsmo~films grown on 9 uc \sto~buffered \ngo~with 2 uc \sto~capping layers. The red dotted line indicates the IR limit for the onset of strong localization \cite{Lee1985}. (b) The corresponding log-plot of resistivity as a function of 1/$T$ respectively. The insets show the low-$T$ zoom-in plot of (I) log($\rho$) versus 1/$T$ for 2 uc, (II) log($\rho$) versus 1/$T^{1/3}$ for 3 uc, and (III) $\rho$ versus log($T$) for the 4 uc film, respectively. Figure reprinted from \cite{Liao2015}.}
\label{ResisNGO}
\end{figure}

We achieved the largest reduction of the electrical dead-layer by growing \lsmo~on \sto-buffered and \sto-capped \ngo~\cite{Liao2015}, see Fig.~\ref{ResisNGO}. In this case a four unit-cell-thick film is still metallic and localization sets in at three unit cells. A few ingredients contribute to the reduction of the dead layer. First of all the films are under compressive strain as opposed to films grown on \sto~that are under tensile strain. Compressive strain induces octahedra rotations around the out-of-plane axis whereas tensile strain induces rotations around the in-plane axes \cite{Vailionis2011}. The double exchange mechanism that drives the conductivity of the manganites depends on the Mn-O-Mn bond angles; strong octahedra rotation reduces the conductivity. Therefore compressive strain promotes double exchange in the critical out-of-plane direction. Second, the adjacent \sto~layers prevent the rotations from the \ngo~from inducing large rotations in the \lsmo~which would be detrimental for the electrical transport. However, it could be expected that the \sto~layers induce octahedra deformation just as is the case of \lsmo~grown on \sto~crystals. Judging from the reduction of the dead layer, this is not so much the case for these heterostructures. Some structural differences are present: the \sto~is strained and probably has an a$^0$a$^0$c$^-$ tilt pattern instead of the standard cubic a$^0$a$^0$a$^0$ tilt pattern and the \lsmo is orthorhombic (a$^+$b$^-$c$^-$ tilt pattern) under compressive strain instead of tetragonal (a$^+$a$^-$c$^0$ tilt pattern) under tensile strain. Presumably one of these structural factors together with the limited thickness of the \sto~layers prevents the \sto~from inducing the Jahn-Teller distorted interfacial layer that occurs in \lsmo~grown on \sto~crystals.

\section{Conclusions}
The magnetic dead-layer at the \lsmo--\sto~interface can be eliminated with the use of compositional interface engineering. The electrical dead-layer, however, remains present in all the heterostructures. The structural distortions induced by oxygen octahedra connectivity are likely the dominant cause for the electrical dead-layer, consistent with the conclusions of a variety of other experiments and theoretical models \cite{Pruneda2007, He2010, Gray2010, Petrov2013}. Also experiments performed on (011) oriented \sto~clearly indicated the importance of the oxygen octahedra rotation pattern on the functional properties \cite{Boschker2012prl}. 

The experiments demonstrate that interface engineering is a powerful tool for studying the properties of materials at interfaces systematically. Due to the importance of the local crystal structure, a main challenge for materials engineers is the control of the amount and direction of the oxygen octahedra rotations at the interface. Next to the results shown in section \ref{liao} other possible approaches to optimize \lsmo~interfaces are using buffer layers with the right rotation patterns, changing the $A$ cation size at the interface, similar to bulk experiments in the manganites \cite{Radaelli1997}, and using infinite layer compounds to eliminate the oxygen octahedra connectivity.

\begin{acknowledgments}
The authors gratefully acknowledge key contributions to the research made by Arturas Vailionis (XRD), Sandra van Aert, Sara Bals, Ricardo Egoavil, Nicolas Gauquelin, Staf van Tendeloo and Jo Verbeeck (STEM), Valeria Lauter, Yaohua Liu and Suzanne te Velthuis (PNR) and Ramamoorthy Ramesh and Pim Rossen (TOF-MS). The transport measurements discussed in section \ref{order} were performed during a stay of H.B. at the University of Tokyo and would not have been possible without the help of Chris Bell, Yasuyuki Hikita, Harold Hwang and Takeaki Yajima.  

\end{acknowledgments}

\bibliographystyle{apsrev4-1}

\end{document}